AN IMPLEMENTATION OF A DUAL-PROCESSOR
SYSTEM ON FPGA

by

Mohammed Eqbal

A Project Presented to the Faculty of the
American University of Sharjah
College of Engineering
in Partial Fulfillment
of the Requirements
for the Degree of

Master of Science in
Computer Engineering

Sharjah, United Arab Emirates

May 2014

## Approval Signatures

We, the undersigned, approve the Master's Project of Mohammed Eqbal.
Project Title: An Implementation of a Dual-Processor System on FPGA.

**Signature**                                                                                            **Date of Signature**
(dd/mm/yyyy)

\_\_\_\_\_\_\_\_\_\_\_\_\_\_\_\_\_\_\_\_\_\_\_\_\_\_\_\_\_\_\_\_\_\_\_\_\_\_\_\_\_\_\_\_\_\_\_\_\_\_\_\_\_\_\_\_\_\_\_\_\_\_\_\_\_\_\_\_\_\_\_\_\_\_\_
Dr. Assim Sagahyroon
Professor, Department of Computer Science and Engineering
Project Advisor

\_\_\_\_\_\_\_\_\_\_\_\_\_\_\_\_\_\_\_\_\_\_\_\_\_\_\_\_\_\_\_\_\_\_\_\_\_\_\_\_\_\_\_\_\_\_\_\_\_\_\_\_\_\_\_\_\_\_\_\_\_\_\_\_\_\_\_\_\_\_\_\_\_\_\_
Dr. Fadi Aloul
Associate Professor, Department of Computer Science and Engineering
Project Co-Advisor

\_\_\_\_\_\_\_\_\_\_\_\_\_\_\_\_\_\_\_\_\_\_\_\_\_\_\_\_\_\_\_\_\_\_\_\_\_\_\_\_\_\_\_\_\_\_\_\_\_\_\_\_\_\_\_\_\_\_\_\_\_\_\_\_\_\_\_\_\_\_\_\_\_\_\_
Dr. Abdul-Rahman Al-Ali
Professor, Department of Computer Science and Engineering
Project Committee Member

\_\_\_\_\_\_\_\_\_\_\_\_\_\_\_\_\_\_\_\_\_\_\_\_\_\_\_\_\_\_\_\_\_\_\_\_\_\_\_\_\_\_\_\_\_\_\_\_\_\_\_\_\_\_\_\_\_\_\_\_\_\_\_\_\_\_\_\_\_\_\_\_\_\_\_
Dr. Tarik Ozkul
Professor, Department of Computer Science and Engineering
Project Committee Member

\_\_\_\_\_\_\_\_\_\_\_\_\_\_\_\_\_\_\_\_\_\_\_\_\_\_\_\_\_\_\_\_\_\_\_\_\_\_\_\_\_\_\_\_\_\_\_\_\_\_\_\_\_\_\_\_\_\_\_\_\_\_\_\_\_\_\_\_\_\_\_\_\_\_\_
Dr. Assim Sagahyroon
Head, Department of Computer Science and Engineering

\_\_\_\_\_\_\_\_\_\_\_\_\_\_\_\_\_\_\_\_\_\_\_\_\_\_\_\_\_\_\_\_\_\_\_\_\_\_\_\_\_\_\_\_\_\_\_\_\_\_\_\_\_\_\_\_\_\_\_\_\_\_\_\_\_\_\_\_\_\_\_\_\_\_\_
Dr. Hany El-Kadi
Associate Dean, College of Engineering

\_\_\_\_\_\_\_\_\_\_\_\_\_\_\_\_\_\_\_\_\_\_\_\_\_\_\_\_\_\_\_\_\_\_\_\_\_\_\_\_\_\_\_\_\_\_\_\_\_\_\_\_\_\_\_\_\_\_\_\_\_\_\_\_\_\_\_\_\_\_\_\_\_\_\_
Dr. Leland Blank
Dean, College of Engineering

\_\_\_\_\_\_\_\_\_\_\_\_\_\_\_\_\_\_\_\_\_\_\_\_\_\_\_\_\_\_\_\_\_\_\_\_\_\_\_\_\_\_\_\_\_\_\_\_\_\_\_\_\_\_\_\_\_\_\_\_\_\_\_\_\_\_\_\_\_\_\_\_\_\_\_
Dr. Khaled Assaleh
Director of Graduate Studies

# Acknowledgment

I have taken much time and effort in this project. However, it would not have been possible without the kind support and help of many individuals. I would like to extend my sincere thanks to all of them.

I am highly indebted to Dr. Assim Sagahyroon for his guidance and constant supervision as well as for providing necessary information regarding the project and also for his support in completing the project.

I would like to express my gratitude towards my professors, especially Dr. Abdul-Rahman Al-Ali, and all the other members of the Computer Engineering Department family for their kind co-operation and encouragement which helped me in the completion of this project.

My thanks and appreciations also go to my family and friends in developing the project who have helped me out with their emotional support.

## Abstract

In recent years, Field-Programmable Gate Arrays (FPGA) have evolved rapidly paving the way for a whole new range of computing paradigms. On the other hand, computer applications are evolving. There is a rising demand for a system that is general-purpose and yet has the processing abilities to accommodate current trends in application processing. This work proposes a design and implementation of a tightly-coupled FPGA-based dual-processor platform. We architect a platform that optimizes the utilization of FPGA resources and allows for the investigation of practical implementation issues such as cache design. The performance of the proposed prototype is then evaluated, as different configurations of a uniprocessor and a dual-processor system are studied and compared against each other and against published results for common industry-standard CPU platforms. The proposed implementation utilizes the Nios II 32-bit embedded soft-core processor architecture designed for the Altera Cyclone III family of FPGAs.

**Search Terms:** FPGA, uniprocessor system, dual-processor system, Nios II processor, Dhrystone.

# Table of Contents

# List of Figures

# List of Tables

# Chapter 1: Introduction

Computer applications are evolving over time. Multimedia applications have lately become a dominant workload on general desktops and workstations, and they are becoming the modern trend [1]. Contemporary applications are rich with multimedia contents. With the explosion of the World-Wide Web and the Internet, workloads in the future are expected to be even more dominated by multimedia [2]. The high degrees of parallelism and large volumes of data, for instance, within such applications have been well-researched, but less well-understood is how to design a processor that is capable of accommodating these intensive applications [3]. Such applications exhibit slightly different characteristics and behavior compared to general applications.

## 1.1 Problem Statement

Traditional superscalar architectures are poorly suited to the demands of modern applications [4]. To enhance the processing of such new trends in applications, engineers have been designing systems that consist of general multi-purpose processors *augmented* with another Application-Specific Integrated Circuit (AISC) which is typically a Graphics Processing Unit (GPU) that assists in the execution of applications. In the context of embedded systems where efficiency is highly demanded, augmenting the main CPU with a GPU is less desired however. Having the two units on separate chips needs a bus in between for communications which impedes performance in terms of power consumption and execution time. Buses usually run at much-slower speeds (clock frequencies) than the CPU clocks. Buses also require more power to transfer information.

There is a growing demand for a system that is general-purpose and yet has the processing capabilities to accommodate the modern trends in application processing. Computer architects, therefore, had to respond to this need by modifying their architectures and evaluating each against the other, in terms of handling contemporary applications. In this work, we propose the design of a Dual-processor System using Field-Programmable Gate Array (FPGA) technology.

### 1.1.1 Characteristics of modern applications.

We have established that modern applications are rich with multimedia contents, and in the near future "workloads are believed to be even more multimedia dominant" [2]. In order to be able to propose or design a computer platform that is general-purpose and yet suits the needs of modern applications, we will initially overview the inherent characteristics of such applications and highlight the differences from other general-purpose applications. Understanding this will enable us to propose a practical dual-processor design.

A very thorough study of multimedia applications' behavior was conducted in Princeton University by Jason Fritts and his colleagues [3]. In their study, Fritts and his colleagues studied a multimedia benchmark called 'MediaBench suite' that is used as an industry standard benchmark to evaluate both compilers and processors. The benchmarks were ranging from image and video processing to audio and speech processing, and even encryption and computer graphics. The compiler used in the study was the IMPACT compiler [3], which allows for extracting and recording the execution behavior. A summary of the results are discussed below.

The study shows that the operation frequencies profile for multimedia is relatively close to the operation frequencies profile of general-purpose applications. Slight differences exist, however, such as the fact that "there is little overall floating point usage", compared to *integer* operations [3]. Besides, integer transfers (load and store) occur more frequently. Therefore, we can conclude that multimedia applications are very memory-intensive applications.

To further understand multimedia applications, few other results have been reported [4], [5]. They concluded that multimedia applications normally had large data sets and seemed to have little data-reuse. Spatial locality was a clear characteristic for these applications, and yet they demonstrated very little global temporal locality. The researchers in [5] also argued that that *cache design* is more important than main memory design, while the study in [4] claimed that multimedia applications have a high degree of inherent *parallelism*. This large amount of parallelism can be exploited on parallel processors.

In Section 1.2, we elaborate more on the FPGA technology.

**1.2  Background Information on FPGA Technology**

FPGA is defined as Field-Programmable Gate Array. It is "a semiconductor device that can be programmed after manufacturing. Instead of being restricted to any predetermined hardware function, an FPGA allows you to program product features and functions, adapt to new standards, and reconfigure hardware for specific applications even after the product has been installed in the field—hence the name 'field-programmable'. You can use an FPGA to implement any logical function that an Application-Specific Integrated Circuit (ASIC) could perform" [6]. Basically, "every FPGA chip is made up of a finite number of predefined resources with programmable interconnects to implement a reconfigurable digital circuit" [7].

FPGA technology allows us to design virtually any computer system in software and then download it to the FPGA chip. This chip then functions according to the design specified using the software as if it was a custom-made ASIC. This technology allows researchers to come up with their own system design, write it as software (in a Hardware Description Language-HDL), download it to the chip, and then test the hardware design. FPGAs are especially suited for industrial and embedded applications as they are "very powerful, relatively inexpensive, and adaptable, since their configuration is specified in an abstract hardware description language" [8].

### 1.2.1  The altera family of FPGAs.

The platform that we are going to use is manufactured by Altera [9]. Altera is a major manufacturer of high-end Programmable Logic Devices (PLD), and is a leading company in the field of programmable logic solutions. The platform solution provided by Altera allows semiconductor companies to quickly and cost-effectively design and build their systems. The company offers FPGAs, System-on-Chips (SoCs) with embedded processor systems, and Complex Programmable Logic Devices (CPLD) in combination with software tools.

### 1.2.2 The cyclone FPGA.

Cyclone III is a member of the Cyclone FPGA family manufactured by Altera. Altera claims that it offers a combination of low power, high functionality, and low cost. Cyclone III FPGA is a good choice for testing our design, as it has features such as "200K logic elements, 8 Mbits of embedded memory, and 396 embedded multipliers". Altera claims that Cyclone III FPGAs are ideal for Video and image processing [10]. Building a Dual-processor System-on-Programmable-Chip can be quite a challenge. We have also selected the Cyclone III FPGA for the comprehensiveness of the documentation available that supports working in a multiprocessor environment.

### 1.2.3 The quratus ii platform.

The system will be implemented using Quartus II platform. Quartus II platform consists of an FPGA Cyclone board complemented with a Quartus II software environment. Quartus II software is a Computer Aided Design (CAD) environment which is basically a tool that enables designing systems on the FPGAs. It is purposely designed to work with the entire range of the FPGA family made by Altera.

### 1.2.4 The SoPC builder.

Part of the Quartus II software environment for FPGAs is a tool called System-on-Programmable-Chip Builder (SoPC Builder). Developers using the SoPC Builder tool can design and build multiprocessor systems. Systems that incorporate two or more processors working together are referred to as multiprocessor systems, and when implemented as a system on a *single* chip they are commonly referred to as Multiprocessor System-on-Chip (MPSoC). Similarly, 'tightly-coupled' multiprocessors refer to systems that have all processors are residing on the same physical chip.

SoPC Builder is a system development tool for creating SoPC and MPSoC design systems by using components such as processors, peripherals, and memories as building blocks [11]. It automates the task of generating and integrating such hardware components. Traditionally, designers must first write the code in Hardware Depiction Language (HDL) for each hardware component, and then manually write other HDL modules to connect together all these components in the system. However, in SoPC Builder, designers can select the system components (as it incorporates a library of components such as the Nios II soft processor, memory controllers, interfaces, and peripherals), configure each component in depth, and then SoPC Builder generates the HDL files required for the components and the interconnect logics automatically. Interconnections are made though the Avalon bus. SoPC Builder generates either Verilog HDL or VHDL equally [12].

### 1.2.5 Soft-hardware components.

The term 'soft-hardware components' refers to hardware-equivalent components that are programmed on the FPGA. Soft-hardware components are first designed in software in HDL, and then downloaded on FPGAs to act like typical hardware. According to [13], "a soft-core microprocessor is a Hardware Description Language (HDL) model of a specific processor (CPU) that can be customized for a given application and synthesized for an ASIC or FPGA target. In many applications, soft-core processors provide several advantages over custom designed processors such as reduced cost, flexibility, platform independence and greater immunity to obsolescence. Embedded systems are hardware and software components working together to perform a specific function".

### 1.2.6 The nios ii processor.

Nios II is a 32-bit embedded soft-core processor architecture designed by Altera. Nios II includes many improvements over the original (16-bit) Nios architecture. According to [14], the Nios II processor can be illustrated as shown in Figure 1.

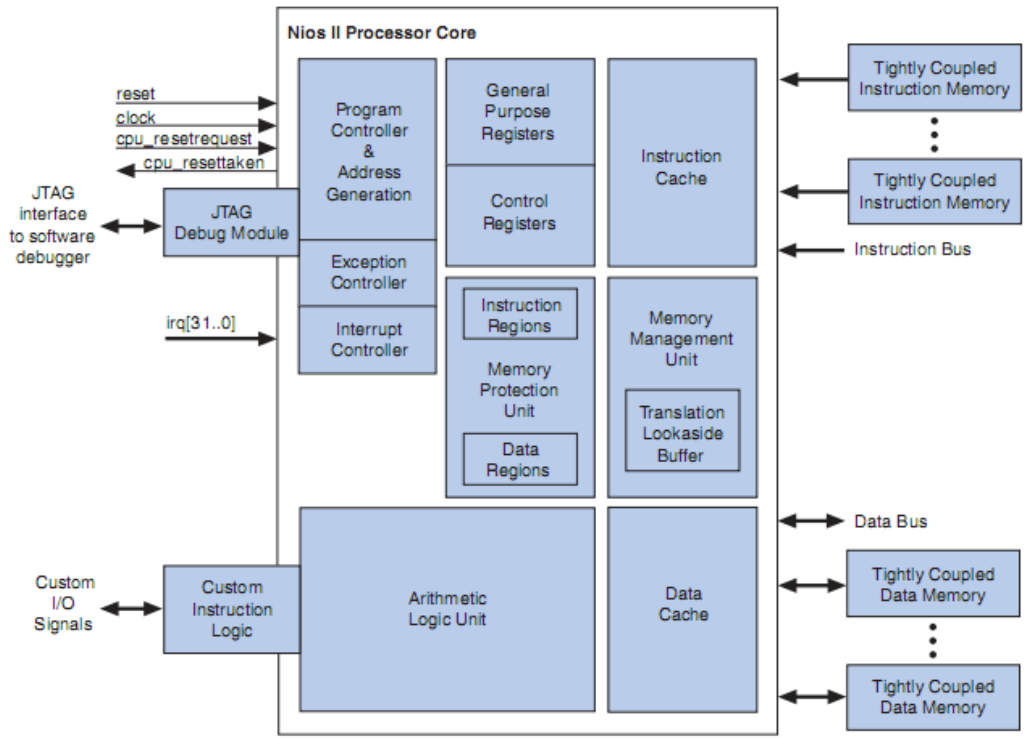

**Figure 1: A Typical Nios II Processor Block Diagram [14]**

According to [15], "the Nios II processor has a Reduced Instruction Set Computer (RISC) architecture. Its arithmetic and logic operations are performed on operands in the general purpose registers. The data is moved between the memory and these registers by means of Load and Store instructions. The word length of the Nios II processor is 32 bits. All registers are 32 bits long. Byte addresses in a 32-bit word can be assigned in either little-endian or big-endian style. The assignment style is one of the options that the user may select at configuration time. The Nios II architecture uses separate instruction and data buses, which is often referred to as the Harvard architecture".

## 1.3 Literature Review and Related Work

In the context of implementing a tightly-coupled dual-processor system on an FPGA that is tailored for modern applications, we discuss some of the work that has been completed and highlight the differences when compared to our approach.

### 1.3.1 Multiprocessor systems on multiple chips.

To achieve higher performance, engineers sometimes utilize heterogeneous multiprocessor platforms which are tuned for a certain well-defined application domain. FPGA is known for providing designers with several benefits in hardware system design customization. In this type of design, the processors are usually laid on different physical chips. The work in [16] discusses the design of an FPGA-based heterogeneous multiprocessor system consisting of four Nios II soft-cores and one ARM core. The ARM core is the central controller of the whole system, and four Nios II cores serve as slaves under control by the ARM core. The ARM core and Nios II cores cooperate and work in parallel to accomplish each task. The System was implemented on Altera Stratix II. Having one central processor governing the Nios soft-core is one approach. In our design, however, we attempt to avoid this by having the dual-processor system on the same chip and thus eradicating any external bus bottleneck possibilities.

### 1.3.2 Multiprocessor system-on-chip (MPSoC).

When creating a multiprocessor system, one approach is to lay multiple processors on different chips (as explored earlier). Another approach is to have the processors on the same chip. This increases system efficiency and performance at the expense of the FPGA chip's real-estate [17].

The study in [18] presented an implementation of a 'tightly coupled', cache-coherent Symmetric Multiprocessing (SMP) architecture using a vendor-provided soft-core processor. By definition, according to [19], SMP is a system "where two or more identical processors connect to one another through a shared memory. Each processor has equal access to the shared memory". Tightly-coupled multiprocessing refers to chip-level multiprocessing, where all processors are residing on the same physical chip. According to [19], in tightly-coupled multiprocessing systems "multiple CPUs are connected via a shared bus (or an interconnect fabric) to a shared memory. Each processor also has its own fast memory. The tightly-coupled nature such systems allows very short physical distances between processors and memory and, therefore, minimal memory access latency and higher performance. This type of architecture works well in multi-threaded applications where threads can be distributed across the processors to operate in parallel". The work in [18] provided a framework for designing an SMP system. Vendor-provided soft-core processors (such as Altera's Nios II) often support advanced features such as caching that work well in uniprocessor or multiprocessor architectures. However, it is a challenge to implement Symmetric Multiprocessor System-on-Programmable-Chip (SMP-SoPC). Given the scope of our project, we will not attempt to provide a framework for Symmetric Multiprocessing (SMP) systems but rather focus on the design aspects of building a tightly-coupled multiprocessor system. Furthermore, the paper in [18] presented results on cache coherency and FPGA utilization. However, it didn't present benchmark results. In our work, we will measure the performance of our implemented system by running a benchmark on it.

A further attempt was made to provide a solid foundation for SMP parallel processing. The researchers in [20] proposed a system called MultiFlex which is basically "an application-to-platform mapping tool that integrates heterogeneous parallel components H/W or S/W into a homogeneous platform programming environment". In our project, we will not attempt to provide a framework for running applications parallel for SMP systems but rather focus on the design aspects of building a tightly-coupled multiprocessor system and its performance. Furthermore in [20] after proposing the mapping tool, an actual Multiprocessor System-on-Chip (MPSoC) was implemented to handle networking and multimedia applications. Testing this proposed system was done by running an internet traffic management application and an MPEG4 video encoder. This work measured the performance in terms of processor utilization percentage which ranged between 85%-91%.

In another work by Tseng and Chen [21] however, an attempt to benchmark the CPU was made. An implementation of an FPGA Nios II based MPSoC was presented and tested. They ran two simple benchmarks that they developed. The first benchmark (VAR) was basically an array that required all processors to operate on the same set of data simultaneously. The second benchmark (ARRAY) was computations for a huge array that was segregated in advance, and distributed to all processors. Processors in the second case were competing against each other only to grab the hardware mutex for accessing the same memory. The researchers designed a four-processor system using Nios II soft-core and implemented an MPSoC architecture. The results showed that "the quad-core system architecture that they proposed can execute the program concurrently at the same time". No industry-standard benchmark was used. Interestingly enough though, they stated that in future work researchers should aim at how to utilize the characteristics of multiprocessor to design multimedia applications. While the work in [21] made a genuine attempt to assess the parallel capabilities of MPSoCs, in our approach we would attempt to run industry-standard benchmark in order to be able to compare our system to other published results, rather than developing the benchmark ourselves.

The work in [22] made a better attempt to evaluate a multiprocessor FPGA implementation. The authors first designed a configurable and scalable multiprocessor system on Xilinx Virtex-5 FPGA (a competitor of the Altera Cyclone FPGA). Then, the performance of a uniprocessor system and a multiprocessor implementation of widely used Fast Fourier Transform (FFT) functions was analyzed, as this was the main objective of the work. On the other hand, this work also presented benchmark results for their implemented systems using CoreMark benchmark which is a part of the EEMBC benchmark suite. The study also concludes that the benchmark "score increases linearly with the number of processors in a multiprocessor system".

While in some work (above) attempts to design, build, and test Multiprocessor System-on-Chip (MPSoC) implementations were made, others only provided a process design flow for building SoC-based embedded systems [17], [23]. Furthermore, other researches considered it enough to simply present an overview of FPGA-based multiprocessor systems, describe the main characteristics, and comment on several FPGA-based multiprocessor systems appearing in the research community for the last five years [24].

When developing a multiprocessor system using Altera's platforms, developers can choose among several memory architectures and configurations of Multiprocessor systems (e.g. shared buffer only and shared memory only, shared nothing …etc.). Sections 1.4 and 2.1 elaborate more on the different architectures supported for designing multiprocessor systems, provide an overview of creating one, and propose a synopsis of our prototype.

## 1.4 Work Outline

Our work in this project can be outlined as follows. After conducting a literature survey and reviewing work that was relevant to ours, we propose a multiprocessor-based system that is designed with the aim of speeding up the execution of applications. We propose a prototype for a uniprocessor and a dual-processor system, using Altera's Quartus II platform and based on the Nios II soft-core processor architecture.

Subsequently, we test the performance of the proposed prototypes. We study different configurations of the uniprocessor and dual-processor system and compare them against each other and against published results for common industry-standard CPU platforms.

# Chapter 2: Proposed System Design and Testing Strategy

## 2.1  Design Overview and System Components

We exhibit the following system diagrams showing the components used in our uniprocessor system (Figure 2) and dual-processor system (Figure 3). These components can be designed and implemented and connected using the 'SoPC Builder' environment.

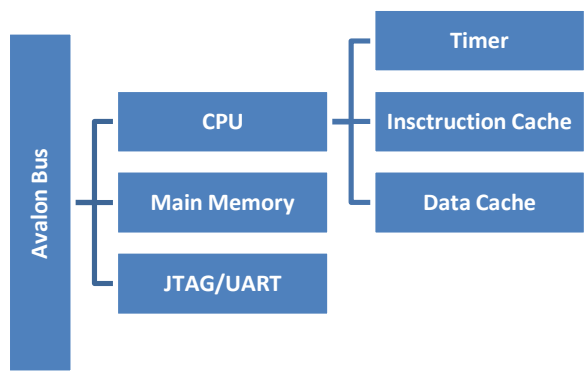

**Figure 2: Components of a Uniprocessor System**

Following is a brief description for the components shown in Figure 2 and Figure 3.

- Each CPU is a Nios II soft-core which consist of a six-stage pipeline running at 66.5 MHz.

- Each CPU has a timer.

- Each CPU has its own separate instruction and data caches. As it was demonstrated (in the Literature Review and Related Work section), heavy emphasis on *cache* was made.

- Main Memory is physically shared for the two CPUs. However, each CPU is assigned to a different segment of the memory.

- Mailbox is the mechanism for the dual-processors to exchange messages. These messages are stored in linked-list structure inside the buffer.

- Mailbox buffer is implemented using the on-chip memory. The on-chip memory can be configured either as the main memory for the CPU, or as buffers for the mailboxes for synchronization. In our case, part of the on-chip memory will be configured as a buffer for the mailbox.

- Mailbox Mutex is a hardware module that governs and guarantees synchronization for accessing the mailbox.

- Avalon bus is the cross-connect fabric that is capable of interconnecting the different components in the system. It is a reconfigurable switch fabric that can interface embedded peripherals and system components.

- JTAG and UART are modules for downloading the design onto the Field-Programmable Gate Array (FPGA) and are also used for debugging.

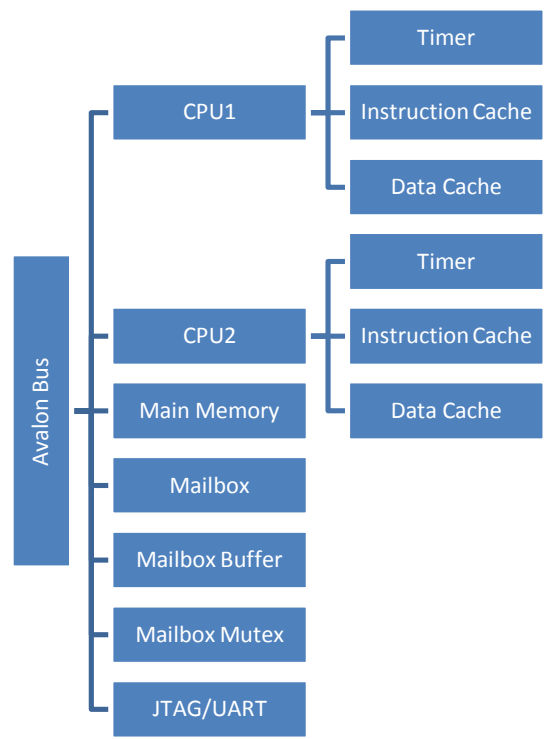

**Figure 3: Components of a Dual-processor System**

Given all the established background information, we can draw the following conclusions that engineers should consider when designing a system that suits modern applications: *Cache* size is heavily emphasized upon and must be well designed and thoroughly tested, and *parallelism* is an also important aspect. We have already said that cache and especially data cache is very crucial. The bigger the data cache size, the better the performance. Furthermore, newer applications require a very high degree of parallelism, having a dual-processor system should, therefore, increase the performance. Additionally, having all hardware components on the same chip (tightly-coupled) must yield efficiency and better performance.

**2.2 Testing Strategy**

We will first build a uniprocessor system and test it with the benchmark. Next, we will build a dual-processor system and similarly examine it, and then compare both.

We will also build different systems, with different instruction cache sizes and different data cache sizes. We will examine how varying any of these aspects improves the performance and produces different results.

**2.3 Design Trade-off**

Furthermore, when designing FPGAs, like any other computer system, an important factor is chip real-estate. When building a dual-processor system, for example, cache sizes are limited to a certain size limit because the FPGA has to simply accommodate more CPU cores which comprise of high-speed registers. Cache also is a very high speed memory. Both CPU cores and cache are implemented on the FPGA using fast M9K memory blocks. In Altera FPGAs, "the embedded memory structure consists of columns of M9K memory blocks". These fast memory blocks can be configured "to provide various memory functions, such as RAM, shift registers, ROM, and FIFO buffers" [25]. In our analysis, we will attempt to answer the following question: To achieve the maximum performance which component is more important and more useful for investing with the given real-estate, cache or the number of CPUs?

## 2.4 Benchmark Selection

We considered 'Dhrystone' benchmark for testing our prototype [26]. Dhrystone is a benchmark program developed in by Reinhold Weicker. Dhrystone grew to become the industry-standard for assessing processors in terms of their performances. Weicker designed Dhrystone by gathering meta-data from a wide range of softwares. Then, he characterized these programs in terms of various common aspects such as procedure calls, pointer indirections, assignments, etc. From this he managed to build the Dhrystone benchmark [27]. One important characteristic of the Dhrystone benchmark is that it is a strong indicator of general-purpose CPU's *integer* performance. A strong character of multimedia application, as Dhrystone benchmark contains no floating point operations.

According to [28], "the C version of Dhrystone is the one mainly used in industry. The original Dhrystone benchmark is still used to measure CPU performance *today*. The original intent with Dhrystone was to create a short benchmark program that was representative of system (integer) programming. The Dhrystone code is dominated by simple integer arithmetic, string operations, logic decisions, and memory accesses intended to reflect the CPU activities in most general purpose computing applications. The Dhrystone result is determined by measuring the average time a processor takes to perform many iterations of a single loop containing a fixed sequence of instructions that make up the benchmark. When Dhrystone is referenced, it is usually quoted as 'DMIPS', or Dhrystone MIPS/MHz". Furthermore, details on the Dhrystone benchmark can further be examined in Appendix A.

The most two popular versions of Dhrystone benchmark are version 1.1 and version 2.1. The difference between the two versions is that the second generation was developed to prevent compiler optimization and provide a true representation of the hardware being tested. In this work, we will use versions 1.1 and 2.1.

## Chapter 3: Results and Discussion

As discussed earlier, we attempt to address the following factors as to how they affect the performance of our system: instruction cache, data cache, and parallelism.

### 3.1 Experiment 1

In this experiment, we build and run different configurations of the system. To be able to study one factor at a time, we remove data cache temporarily from our design in this experiment, and build several systems while varying the *instruction cache size*. The following instruction cache sizes are used: 2 KB, 4 KB, 8 KB, and 16 KB, when building *uniprocessor* systems. We also use the same instruction cache sizes set for *dual-processor* system this time to observe the difference. It is important to note that while studying a certain aspect of the design (such as instruction cache) other factors are fixed. In Table 1, each row represents a certain configuration of the system being tested.

Table 1: Instruction Cache and Number of CPU Analysis

| Number of CPUs | IC Size | DC Size | Dhrystones 1.1/Second | Dhrystones 2.1/Second |
|---|---|---|---|---|
| 1 | 2 KB | 0 | 10000 | 10204 |
| 1 | 4 KB | 0 | 10638 | 10537 |
| 1 | 8 KB | 0 | 11238 | 10870 |
| 1 | 16 KB | 0 | 11297 | 11364 |
| 2 | 2 KB | 0 | 19999 | 20409 |
| 2 | 4 KB | 0 | 21277 | 21074 |
| 2 | 8 KB | 0 | 22471 | 21738 |
| 2 | 16 KB | 0 | 22473 | 22727 |

The first column (in Table 1) indicates the number of CPUs used in each configuration. The second column indicates the instruction cache size (in KB) that is used. The third column shows that data cache is not implemented in this experiment. The forth and the fifth columns show the benchmark results for Dhrystone versions 1.1 and 2.1, respectively, recorded in Dhrystones per seconds for the given hardware configuration.

From the results shown in Table 1, we first observe that *introducing a second CPU yields a speedup factor of around 2* consistently and throughout all instruction cache size variations.

Figure 4 shows a plot of the benchmark results, over the different configurations of the system that are implemented. On the y-axis, the values are in Dhrystones per seconds. The blue curve represents the benchmark results for Dhrystone version 1.1, while the red curve represents the benchmark results for Dhrystone version 2.1. On the x-axis, we start by having one CPU (uniprocessor system) with an instruction cache of size 2 KB. Then, we increase the instruction cache gradually by using the following sizes for the instruction cache: 4 KB, 8 KB, and 16 KB. Next, we add a second CPU (dual-processor system), and set the instruction cache size back to 2 KB. Similarly, we increase the instruction cache gradually by using the following sizes for the instruction cache: 4 KB, 8 KB, and 16 KB.

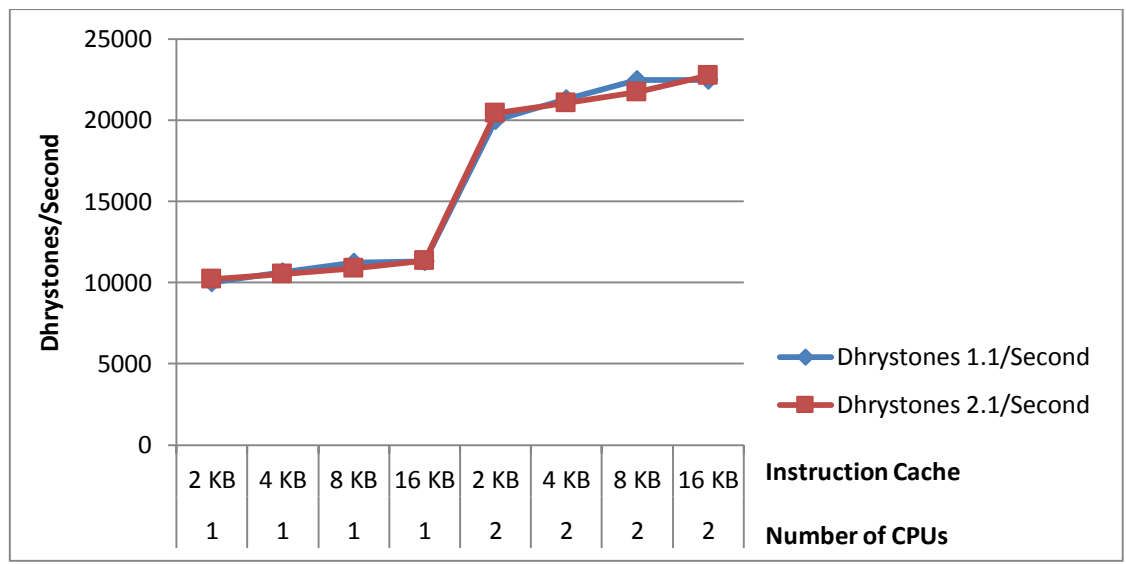

**Figure 4: Instruction Cache and Number of CPU Analysis**

The Feld-Programmable Gate Array (FPGA) resources does not allow to increase the instruction cache size any further beyond 16 KB, as we will discuss in the 'Resource usage' section.

As seen in Figure 4, if we consider the first set of four configurations (when the number of CPUs is 1), we notice that *increasing the instruction cache increases the performance only slightly*. The same can be observed with the second set of four configurations (when the number of CPU is 2).

### 3.1.1 Resource usage.

Referring to the last row in Table 1, it is important to note that the Cyclone III FPGA simply cannot fit two CPUs with more than 16 KB of instruction cache. Table 2 shows the FPGA hardware usage summary for this particular configuration. Note that for the M9K blocks (defined in Section 2.3), 51 blocks are already used out of the 66 available blocks. If we increase the instruction cache to more than 16 KB the design simply will not fit the FPGA anymore.

In Table 2, the first column lists the different hardware resources that are available on the FPGA. The second column shows, for the given resource, how many are used out of the total number of available resources on the FPGA in this system configuration. This is also represented as a percentage (in parentheses) in the second column.

Table 2: Resource Usage for IC 16 KB, 2 CPUs

| Resource | Usage |
| --- | --- |
| Total logic elements | 9,718 / 24,624 ( 39 % ) |
| Total registers | 5,460 / 25,629 ( 21 % ) |
| Total LABs: partially or completely used | 790 / 1,539 ( 51 % ) |
| I/O pins | 131 / 216 ( 61 % ) |
| **M9Ks** | **51 / 66 ( 77 % )** |
| Total block memory bits | 338,592 / 608,256 ( 56 % ) |
| Total block memory implementation bits | 470,016 / 608,256 ( 77 % ) |
| Embedded Multiplier 9-bit elements | 8 / 132 ( 6 % ) |
| PLLs | 1 / 4 ( 25 % ) |
| Global clocks | 20 / 20 ( 100 % ) |

| JTAGs | 1 / 1 ( 100 % ) |
|---|---|
| CRC blocks | 0 / 1 ( 0 % ) |
| ASMI blocks | 0 / 1 ( 0 % ) |
| Impedance control blocks | 0 / 4 ( 0 % ) |

From the results presented in Table 2, we conclude the following: Increasing the instruction cache size increased the performance slightly, while introducing a second CPU doubled the performance.

**3.2 Experiment 2**

In this experiment, we attempt to study the effect of varying the *data cache size*. Hence, we fix the instruction cache size at 8 KB. We also vary the *number of CPUs* used. Table 3 shows the results of this experiment.

Table 3: Data Cache and Number of CPU Analysis

| Number of CPUs | IC Size | DC Size | Dhrystones 1.1/Second | Dhrystones 2.1/Second |
|---|---|---|---|---|
| 1 | 8 KB | 4 KB | 41666 | 41667 |
| 1 | 8 KB | 8 KB | 41670 | 41667 |
| 1 | 8 KB | 16 KB | 45450 | 41667 |
| 1 | 8 KB | 32 KB | 45454 | 41667 |
| 2 | 8 KB | 4 KB | 83332 | 68234 |
| 2 | 8 KB | 8 KB | 87118 | 71428 |

In Table 3, the first column shows the number of CPUs implemented. The second column indicates the instruction cache size (in KB) that was used. The third column indicates the data cache size (in KB) that was used. The forth and the fifth columns show the benchmark results for Dhrystone versions 1.1 and 2.1, respectively.

When comparing the results for this experiment (in Table 3) with the results in the previous experiment (in Table 1), the first observation to make is that in the previous experiment the overall maximum number of Dhrystones 1.1 per second that we achieved was only 22473. Similarly, for the Dhrystones 2.1 per second, the maximum number was 22727. Note that the sizes of the instruction cache used in the previous experiment were up to 16 KB. In this experiment however, when we *introduce a data cache of only 4 KB, the benchmark results nearly double*. We also notice that, after the data cache was added, *increasing the data cache size increases the performance slightly*. This can be clearly observed in Figure 5.

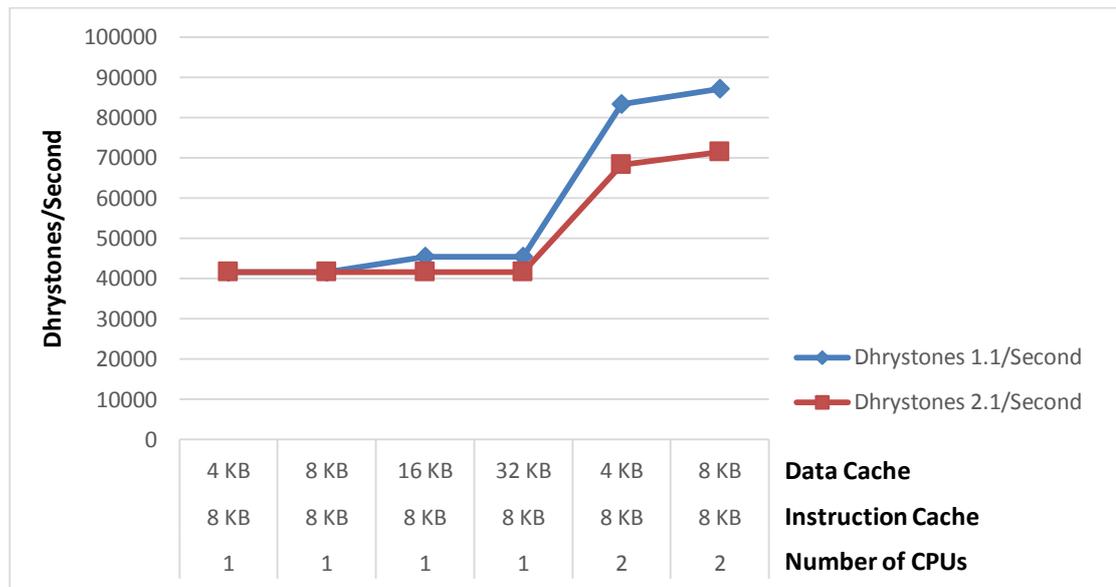

**Figure 5: Data Cache and Number of CPU Analysis**

Figure 5 shows a plot of the benchmark results. On the y-axis, the values are in Dhrystones per seconds. On the x-axis, the first system configuration used is as follows: Number of CPU is one, instruction cache size is 8 KB, and data cache size is 4 KB. We fix the instruction cache size at 8 KB throughout this experiment. We increase the data cache gradually by using the following sizes for the data cache: 4 KB, 8 KB, 16 KB, and 32 KB. The FPGA resources does not allow to increase the data cache size any further. Next, we set the data cache size back to 4 KB and add a second CPU. Similarly, we increase the data cache size to 4 KB, and then to 8 KB. The resources of the FPGA does not allow to increase the data cache size any further.

We observe that (similar to the results obtained in the previous experiment) *marginal performance speedup was gained when we increased the data cache size*. We believe that this is due to the limitation of the benchmark that we are using. A different (larger, in working data set) benchmark would have exploited the larger data cache more effectively and yielded even better results.

Furthermore, if we take the *second* configuration and the *sixth* configuration in Table 3, we notice that for both configurations, the data and instruction cache size was 8 KB, but a second CPU was added to the latter one. We hence conclude that *adding a second CPU nearly doubled the performance*.

### 3.2.1 Resource usage.

To establish the last conclusion to be drawn from this experiment, we assess the resource utilization report for the *fourth* and the *sixth* configuration (in Table 3). For the *fourth* configuration (in Table 3), when the number of CPU is 1 and the instruction cache size is fixed at 8 KB, the *maximum* data cache size that can fit the Cyclone III FPGA is 32 KB. The 'Quratus II Fitter' tool basically cannot accommodate a data cache size of 64 KB for this particular configuration (refer to Table 4 to review the resource utilization report). We can observe from the table that 80% the M9K blocks are already used.

In Table 4, the first column lists the resources available on the FPGA. The second column shows how many resources were used out of the available resources on the FPGA in this particular configuration (1 CPU, 8 KB instruction cache, 32 KB data cache).

**Table 4: Resource Usage for IC 8 KB, DC 32 KB, 1 CPU**

| Resource | Usage |
| --- | --- |
| Total logic elements | 8,937 / 24,624 ( 36 % ) |
| Total registers | 5,492 / 25,629 ( 21 % ) |
| Total LABs: partially or completely used | 682 / 1,539 ( 44 % ) |
| I/O pins | 131 / 216 ( 61 % ) |
| **M9Ks** | 53 / 66 ( 80 % ) |
| Total block memory bits | 374,944 / 608,256 ( 62 % ) |

| Total block memory implementation bits | 488,448 / 608,256 ( 80 % ) |
|---|---|
| Embedded Multiplier 9-bit elements | 4 / 132 ( 3 % ) |
| PLLs | 1 / 4 ( 25 % ) |
| Global clocks | 20 / 20 ( 100 % ) |
| JTAGs | 1 / 1 ( 100 % ) |
| CRC blocks | 0 / 1 ( 0 % ) |
| ASMI blocks | 0 / 1 ( 0 % ) |
| Impedance control blocks | 0 / 4 ( 0 % ) |

Similarly for the *sixth* configuration (the last row in Table 3), the utilization report is shown in Table 5.

Table 5: Resource Usage for IC 8 KB, DC 8 KB, 2 CPUs

| Resource | Usage |
|---|---|
| Total logic elements | 11,813 / 24,624 ( 48 % ) |
| Total registers | 6,631 / 25,629 ( 26 % ) |
| Total LABs: partially or completely used | 906 / 1,539 ( 59 % ) |
| I/O pins | 131 / 216 ( 61 % ) |
| **M9Ks** | 55 / 66 ( 83 % ) |
| Total block memory bits | 366,752 / 608,256 ( 60 % ) |
| Total block memory implementation bits | 506,880 / 608,256 ( 83 % ) |
| Embedded Multiplier 9-bit elements | 8 / 132 ( 6 % ) |
| PLLs | 1 / 4 ( 25 % ) |
| Global clocks | 20 / 20 ( 100 % ) |
| JTAGs | 1 / 1 ( 100 % ) |
| CRC blocks | 0 / 1 ( 0 % ) |
| ASMI blocks | 0 / 1 ( 0 % ) |
| Impedance control blocks | 0 / 4 ( 0 % ) |

In Table 5, the first column lists the different hardware resources that are available. The second column shows how many resources are used out of the available resources on the FPGA in this particular configuration (2 CPU, 8 KB instruction cache, 8 KB data cache). Note the number of M9K blocks used. Having a larger data cache sizes with 2 CPUs fitted on the same FPGA is not physically possible.

A very important factor is to be considered at this point and a solid conclusion must be drawn. When designing a tightly-coupled system, designers are often confronted with the following challenge: Given the same real estate on the chip, which is more important, number of CPUs or cache size? In other words, to fully utilize the chip resources and optimize the execution should the designer add more cores or should the designer take advantage of the chip area by having bigger cache size? We clearly answer this question by reviewing the results of this experiment: *Larger number of CPUs adds more performance than increasing data cache*.

We can now summarize the findings in this experiment as follows. Having a data cache is very critical for performance. Introducing a data cache in a system nearly doubles the benchmark results. Increasing the data cache size increases the performance slightly. Adding a second CPU nearly doubles the performance. Higher number of CPUs enhances performance more than increasing the data cache size.

## 3.3 Conclusions from the Experiments

From the two experiments above we can summarize all our findings as follows.

- Increasing the instruction cache size increased the performance slightly.
- Adding a data cache, nearly doubles the performance.
- Increasing the data cache size increases the performance only slightly.
- Introducing a second CPU doubles the performance.
- Having more CPU cores enhances the performance more than increasing data cache.

## 3.4 Design Evaluation

By combining all the findings above, we can consider the following system with the following configuration (Table 6) for designing a Dual-processor System on a Cyclone III FPGA. The source code for the proposed dual-processor system can be examined in **Error! Reference source not found.**. This Nios II Dual-processor System configuration can be compared to other industry-standard CPU platforms.

**Table 6: Proposed Nios II Dual-Processor System Design Configuration**

| Number of CPUs: | **2 CPUs** |
|---|---|
| Instruction Cache Size: | **8 KB** |
| Data Cache Size: | **8 KB** |

### 3.4.1 Benchmark results.

The benchmark results for the described system design (in Table 6) are shown in Table 7.

**Table 7: Proposed Nios II Dual-Processor System Benchmark Results**

| **Dhrystone Version or Type** | **Results** |
|---|---|
| Dhrystones 1.1/Second | **87118** |
| Dhrystones 2.1/Second | **71428** |
| VAX MIPS (1.1) | **49.58** |
| VAX MIPS (2.1) | **40.65** |

The first two benchmark results in the first two rows (in Table 7) are recorded in terms of Dhrystones per second. To obtain VAX MIPS results, we convert by dividing Dhrystones per second (for both versions 1.1 and 2.1) by 1757 which is the DEC VAX 11/780 result. According to [29], "the VAX-11 was a family of minicomputers developed and manufactured by Digital Equipment Corporation (DEC). The VAX-11/780 was the first VAX computer". Researchers decided to record the Dhrystone benchmark on this machine and record the result. Then, other machines' results must be divided by the result of this machine (the VAX-11/780). This would give us the ratio between the performances of the machines being tested to the performance of the VAX-11/780 which was 1757 Dhrystones/Seconds.

## 3.5 Comparison with Industry-Standard Benchmark Results

Table 8 and Table 9 show benchmark results of Dhrystone 1.1 and 2.1 respectively, for several industry-standard CPUs, including our Proposed Nios II Dual-Processor System [30]. The first column shows the name of the CPU. The second column represents the clock speed (frequency, in MHz) that the CPU runs at. The third column shows the Dhrystone benchmark results in VAX MIPS format.

Table 8: VAX MIPS 1.1 Benchmark Results on PCs

| CPU | MHz | VAX MIPS 1.1 |
|---|---|---|
| AMD 80386 | 40 | 4.32 |
| IBM 486D2 | 50 | 7.89 |
| AMD 5X86 | 133 | 9.37 |
| IBM 486BL | 100 | 12 |
| 80486 DX2 | 66 | 12 |
| **Nios II Dual-Processor System** | **66.5** | **49.58** |
| AMD K62 | 500 | 77.8 |
| AMD K63 | 450 | 76.3 |

Table 9: VAX MIPS 2.1 Benchmark Results on PCs

| CPU | MHz | VAX MIPS 2.1 |
|---|---|---|
| AMD 80386 | 40 | 4.53 |
| IBM 486D2 | 50 | 7.89 |
| AMD 5X86 | 133 | 9.42 |
| IBM 486BL | 100 | 11.8 |
| 80486 DX2 | 66 | 12.4 |
| **Nios II Dual-Processor System** | **66.5** | **40.65** |
| AMD K6 | 200 | 43.3 |
| IBM 6x86 | 150 | 43.9 |

Note that our proposed Nios II Dual-Processor System's performance (that is running at 66.5 MHz) is better than the 80486 DX2, according to the results demonstrated in Table 8 and in Table 9.

# Chapter 4: Conclusion

Field-Programmable Gate Arrays (FPGA) are excellent platforms to test and evaluate computer hardware architectures. Tightly-coupled Multiprocessor systems eliminate bottle-necks and provide a better performance than multiprocessor systems on multiple chips (loosely-coupled systems).

In multiprocessor systems, increasing the *instruction cache* size increases the performance slightly. Moreover, having a *data cache* is very critical for performance, as adding a data cache to the design nearly doubles the performance. After adding the data cache, increasing its size also increases the performance slightly. As for the *number of CPU* cores in the system, adding a second CPU doubles the performance. Increasing the number of CPUs increases the performance linearly.

When designing a tightly-coupled system, engineers are often confronted with the following challenge: Given the same real estate on the chip, which is more important, number of CPUs or cache size? Based on this work we conclude that having more CPUs enhances performance more than increasing data cache.

# Chapter 5: Future Work and Recommendations

As an expansion for this work the following aspects can be considered.

We ran the design on a Cyclone III chip. Yet, newer versions of Cyclone family are available. However, the issue with newer platforms is that as Altera is keen on rolling out new versions of their FPGAs into the market to maintain a competitive edge, support from both Altera and the developers community remains poor for newer models, especially for designing multiprocessor platforms. At the moment, designing a multiprocessor system on Cyclone III remains very well established in this regard. Nonetheless, for future expansions new versions must be considered.

The Nios II was running at 66.5 MHz, which is one of the available frequencies on the educational board. The FPGA can run on higher frequencies if faster clocks are available.

Researchers must consider running benchmarks from the Embedded Microprocessor Benchmark Consortium (EEMBC) on the design configuration of Nios II Dual-Processor System that we proposed. EEMBC is a well-established organization that is specialized in systems benchmarking and publishing benchmark results. All EEMBC benchmarks are licensed and need to be purchased before using [31]. Given the resources allocated for this project, we were not able to purchase the license.

It is worth mentioning that before selecting Dhrystone as the benchmark to test our prototype, we examined the following benchmarks: ALPBench [32], Berkeley multimedia workload [33], MiBench [34], OpenMPBench [35]. The issue with the first three mentioned benchmarks is that they all are based on the concept of receiving 'files' as an input and producing an output in the form of 'files' as well. Most embedded systems do not support (and especially Nios II) a file system. As a workaround for this issue, (using the debugging software) the input files can be 'fed' into the Nios II's memory as a 'stream of bytes', one by one. The downside of this approach is that this will hinder the performance dramatically and the benchmark results will not truly reflect the capability of the system hardware being tested. Furthermore, there is an issue that we faced with the ALPBench and Barklay workloads. The issue was that the source code for both benchmarks contained 'in-line assembly' code. Assembly language is CPU-architecture-specific, and running it on Nios II requires practically re-writing the whole benchmark core.

In summary, all the benchmarks mentioned above are simple academic projects that are *mere attempts* to create acceptable benchmarks for embedded systems. This shows that there is great demand for such applications and room for research in this area.

# Appendix A

This section provides a detailed description for the Dhrystone benchmark version 2.1 [36], as well as the driver program for the dual-processor system.

```
The following program contains statements of a high level programming
language (here: C) in a distribution considered representative:

  assignments              52 (51.0 %)
  control statements       33 (32.4 %)
  procedure, function calls 17 (16.7 %)

103 statements are dynamically executed. The program is balanced with
respect to the three aspects:

  - statement type
  - operand type
  - operand locality
       operand global, local, parameter, or constant.

The combination of these three aspects is balanced only approximately.

1. Statement Type:
-----------------           number

  V1 = V2                     9
     (incl. V1 = F(..)
  V = Constant               12
  Assignment,                 7
    with array element
  Assignment,                 6
    with record component
                             --
                             34      34

  X = Y +|-|"&&"|"|" Z        5
  X = Y +|-|"==" Constant     6
  X = X +|- 1                 3
  X = Y *|/ Z                 2
  X = Expression,             1
        two operators
  X = Expression,             1
        three operators
                             --
                             18      18

  if ....                    14
    with "else"      7
    without "else"   7
        executed         3
        not executed     4
  for ...                     7 | counted every time
  while ...                   4 | the loop condition
  do ... while                1 | is evaluated
  switch ...                  1
  break                       1
  declaration with            1
    initialization
                             --
                             34      34

  P (...)  procedure call    11
    user procedure      10
    library procedure    1
  X = F (...)
        function   call       6
    user function       5
    library function    1
                             --
                             17      17
```

```
                                   ---
                                   103
```

The average number of parameters in procedure or function calls
is 1.82

2. Operators
------------

|                  | number | approximate percentage |
|------------------|--------|------------------------|
| Arithmetic       | 32     | 50.8                   |
| +                | 21     | 33.3                   |
| -                | 7      | 11.1                   |
| *                | 3      | 4.8                    |
| / (int div)      | 1      | 1.6                    |
| Comparison       | 27     | 42.8                   |
| ==               | 9      | 14.3                   |
| /=               | 4      | 6.3                    |
| >                | 1      | 1.6                    |
| <                | 3      | 4.8                    |
| >=               | 1      | 1.6                    |
| <=               | 9      | 14.3                   |
| Logic            | 4      | 6.3                    |
| && (AND-THEN)    | 1      | 1.6                    |
| \| (OR)          | 1      | 1.6                    |
| ! (NOT)          | 2      | 3.2                    |
|                  | 63     | 100.1                  |

3. Operand Type (counted once per operand reference):
---------------

|           | number | approximate percentage |
|-----------|--------|------------------------|
| Integer   | 175    | 72.3 %                 |
| Character | 45     | 18.6 %                 |
| Pointer   | 12     | 5.0 %                  |
| String30  | 6      | 2.5 %                  |
| Array     | 2      | 0.8 %                  |
| Record    | 2      | 0.8 %                  |
|           | 242    | 100.0 %                |

When there is an access path leading to the final operand (e.g. a record
component), only the final data type on the access path is counted.

4. Operand Locality:
------------------

|                  | number | approximate percentage |
|------------------|--------|------------------------|
| local variable   | 114    | 47.1 %                 |
| global variable  | 22     | 9.1 %                  |
| parameter        | 45     | 18.6 %                 |
| value            | 23     | 9.5 %                  |
| reference        | 22     | 9.1 %                  |
| function result  | 6      | 2.5 %                  |
| constant         | 55     | 22.7 %                 |
|                  | 242    | 100.0 %                |

The program does not compute anything meaningful, but it is syntactically
and semantically correct. All variables have a value assigned to them
before they are used as a source operand.

There has been no explicit effort to account for the effects of a
cache, or to balance the use of long or short displacements for code or
data.

The following program shows the source code of the driver program for the dual-processor system.

```
/*************************************Main CPU1*************************************/

#include <stdio.h>
#include "nios2.h"
#include "altera_avalon_mailbox.h"
#include "dhrystoneNIOS.h"

int main()
{
        // CPU ID, var. initializations
        int id=0;
        int r2=0;
        int r1=0;
        int error_code = 0;

        // Pointer to our mailbox device
        alt_mailbox_dev* mailbox = NULL;

        // Get our processor ID
        NIOS2_READ_CPUID(id);
        printf ("Hello from CPU%d!\n\r", id);

        /*
         * Open the mailbox which provides mutually exclusive access to the messages
to CPU2.
         */
        mailbox = altera_avalon_mailbox_open("/dev/message_buffer_mailbox");
        //0x0400d514;
        /*
         * Check for successful mailbox open.
         */
        if (mailbox)
        {

                while(1)
                {
                        printf ("CPU%d: Running Dhrystones!\n\r", id);
                        r1=dhrystone();

                    printf ("CPU%d: My Dhrystone result is %d\n", id, r1);
                    //i++;
                    printf ("CPU%d: Waiting to read from CPU2.\n", id);
                    r2=altera_avalon_mailbox_get(mailbox, &error_code); //This function
is non-blocking!

                printf ("CPU%d: CPU2's Dhrystone result is %d\n", id, r2, error_code);
                    }

        }
        else
        {
                        printf ("CPU%d: Cannot open Mailbox!\n\r", id);
        }/* mailbox */

  return 0;
}
```

```
/*************************************Main CPU2************************************/

#include <stdio.h>
#include "nios2.h"
#include "altera_avalon_mailbox.h"
#include "dhrystoneNIOS.h"

int main()
{
        // CPU ID
        int id=0;
        int flag=0;
        int r=0;

        // Pointer to our mailbox device
        alt_mailbox_dev* mailbox = NULL;

        // Get our processor ID
        NIOS2_READ_CPUID(id);
        //printf ("Hello from CPU%d!\n\r", id);

        /*
         * Open the mailbox which provides mutually exclusive access to the messages
to CPU2.
         */
        mailbox = altera_avalon_mailbox_open("/dev/message_buffer_mailbox");

        /*
         * Check for successful mailbox open.
         */
        if (mailbox)
        {
                while(1)
                {
                        //printf ("CPU%d: Running Dhrystones!\n\r", id);
                        r=dhrystone();

                if (altera_avalon_mailbox_post (mailbox, r))
                        {
                            //printf ("CPU%d: Mailbox full! i=%d\n\r", id, r);
                        }
                //i++;
                }
        }
        else
        {
                //printf ("CPU%d: Cannot open Mailbox!\n\r", id);
        }/* mailbox */

  return 0;
}

//PROJECT REPORT VER 16 REDUCED
```

# Vita

Mohammed Eqbal was born in 1985 in Aden, Yemen. He was educated in local public schools, and later travelled to the United Arab Emirates to get his high school education. He graduated from Al-Shola Private School in 2003 in Sharjah, with a grade average of 96.3%. He was given many certificates and awards in each of the three years of high school education. This allowed him to receive a Scholarship from the American University of Sharjah, from where he graduated in 2008 with a Bachelor's degree in Computer Engineering.

Upon graduation, Mr. Eqbal began his Master's education and was offered a Graduate Teaching Assistantship at the American University of Sharjah, where he worked as a Laboratory Assistant for two years.

Mr. Mohammed Eqbal currently works as the Group I.T. Manager at Grand Capital group of companies in Dubai, United Arab Emirates. He is also an active member of the Association for Computer Machinery (ACM) and the IEEE - Computer Society. He has published a paper in 2010 titled "Drawing on the Benefits of RFID and Bluetooth Technologies", at the IEEE Asia Pacific Conference on Circuits and Systems.